\documentclass[11pt]{article}
\usepackage{amssymb,amsmath,amsfonts}
\usepackage{graphicx}
\usepackage{graphics}
\usepackage{eepic,epsfig}

\begin{document}

\title{Vacuum currents in partially compactified Rindler spacetime \\
with an application to cylindrical black holes}
\author{V. Kh. Kotanjyan$^{1}$,\, A. A. Saharian$^{1}$,\, M. R. Setare$^{2}$
\\
\\
\textit{$^1$ Department of Physics, Yerevan State University,}\\
\textit{1 Alex Manoogian Street, 0025 Yerevan, Armenia} \vspace{0.3cm}\\
\textit{$^{2}$ Department of Physics, Campus of Bijar, University of
Kurdistan,} \\
\textit{Bijar, Iran}}
\maketitle

\begin{abstract}
The vacuum expectation value of the current density for a charged scalar
field is investigated in Rindler spacetime with a part of spatial dimensions
compactified to a torus. It is assumed that the field is prepared in the
Fulling-Rindler vacuum state. For general values of the phases in the
periodicity conditions and the lengths of compact dimensions, the
expressions are provided for the Hadamard function and vacuum currents. The
current density along compact dimensions is a periodic function of the
magnetic flux enclosed by those dimensions and vanishes on the Rindler
horizon. The obtained results are compared with the corresponding currents
in the Minkowski vacuum. The near-horizon and large-distance asymptotics are
discussed for the vacuum currents around cylindrical black holes. In the
near-horizon approximation the lengths of compact dimensions are determined
by the horizon radius. At large distances from the horizon the geometry is
approximated by a locally anti-de Sitter spacetime with toroidally compact
dimensions and the lengths of compact dimensions are determined by negative
cosmological constant.
\end{abstract}


\bigskip

\section{Introduction}

In quantum field theory the vacuum state, in general, depends on the choice
of the complete set of mode functions used for the expansion of the field
operator (see, for instance, \cite{Birr82}). Among the most known examples
are the Minkowski and Fulling-Rindler (FR) vacua in Minkowski spacetime. The
first one is realized by the plane-wave modes most frequently used in
considerations of quantum field-theoretical effects on background of flat
spacetime. The FR vacuum \cite{Full73} corresponds to the quantization of
fields in the reference frame with the coordinate lines corresponding to the
worldlines of uniformly accelerated observers (Rindler coordinates). On the
base of the equivalence principle, we can expect that properties of quantum
fluctuations in the FR vacuum will have common qualitative features with
those for vacuum states of quantum fields in classical gravitational
backgrounds. In particular, the Rindler metric is the leading approximation
of the near-horizon geometry for most black holes. A better understanding of
quantum vacuum effects in Rindler spacetime serves as a handle in
considerations of more complicated background geometries like Schwarzschild.
Another motivation for the study of the properties of the FR vacuum is
related to the conformal connection of the Rindler metric tensor to the
metric tensors of de Sitter spacetime and of the Friedmann-Robertson-Walker
spacetime with negative spatial curvature \cite{Birr82}. By using these
conformal relations, the expectation values of local physical observables
for conformally invariant fields in those spacetimes are obtained from the
Rindler expectation values by conformal transformations. Rindler observers
in anti-de Sitter (AdS) spacetime and the related AdS/CFT correspondence
have been discussed in \cite{Empa99}-\cite{Pari18}.

Among the interesting topics in the investigations of the properties for a
given vacuum state is the influence of background topology and of boundaries
having different physical nature. The corresponding quantum
field-theoretical effects are known under the general name Casimir effect
(for reviews see \cite{Most97}-\cite{Casi11}). The influence of uniformly
accelerated planar boundaries on both the local and global characteristics
of the FR vacuum has been studied in \cite{Cand77}-\cite{Saha06}. The bulk
and surface Casimir densities for a spherical boundary in Rindler-like
spacetimes were discussed in \cite{Saha05,Saha06b}. In the present paper we
consider the effects of nontrivial topology on the expectation value of the
current density for a charged scalar field in a locally Rindler spacetime
with a part of spatial dimensions compactified to a torus. In appendix \ref%
{sec:App1} we show that the corresponding metric tensor describes the
near-horizon geometry of cylindrical black holes (black strings in another
terminology, for geometrical properties see \cite{Lemo95}-\cite{Gaet17} and
references therein). The field-theoretical models with compact dimensions
appear in a large number of physical problems including the
Kaluza-Klein-type theories, supergravity and string theories, and effective
field-theoretical models in condensed matter physics (for example, the Dirac
model in the long-wavelength description of graphene nanotubes and
nanoloops). The topological Casimir effect has been considered previously as
a stabilization mechanism in theories with extra spatial dimensions and as a
source of the dark energy driving the accelerated expansion of the universe.
In addition, the vacuum expectation values (VEVs) of the current densities
along compact dimensions may serve as a source for large scale magnetic
fields in models with extra dimensions.

The paper is organized as follows. In the next section we describe the
geometry of the problem and present the complete set of scalar modes. The
properties of the vacuum state are encoded in two-point functions and in
Section \ref{sec:Hadam} we evaluate the Hadamard function for a general
number of compact and uncompact dimensions. By using the expression for the
Hadamard function, the VEV of the current density along compact dimensions
is investigated in Section \ref{sec:Current}. The main results of the paper
are summarized in Section \ref{sec:Conc}. In Appendix \ref{sec:App2}, a
representation of the Hadamard function for the Minkowski vacuum is provided
that is used in the main text for the investigation of the effects induced
by acceleration. In Appendix \ref{sec:App1} we demonstrate that the
near-horizon geometry of cylindrical black holes is approximated by Rindler
spacetime with toral spatial dimensions.

\section{Scalar field modes in Rindler spacetime with compact dimensions}

\label{sec:Modes}

We start by the description of the background geometry. The latter is given
by $(D+1)$-dimensional locally Rindler line element
\begin{equation}
ds^{2}=\xi ^{2}d\tau ^{2}-d\xi ^{2}-\sum_{i=2}^{D}\left( dx^{i}\right) ^{2},
\label{ds2}
\end{equation}%
with the Rindler coordinates $x^{\mu }=\left( x^{0}=\tau ,x^{1}=\xi ,\mathbf{%
x}_{p},\mathbf{x}_{q}\right) $ and $0<\xi <\infty $. The $p$-dimensional
subspace with Cartesian coordinates $\mathbf{x}_{p}=(x^{2},...,x^{p+1})$ has
trivial topology $R^{p}$ with $-\infty <x^{l}<\infty $ for $l=2,..,p+1$. The
$q$-dimensional subspace covered by the coordinates $\mathbf{x}%
_{q}=(x^{p+2},...,x^{D})$ is compactified to a torus $(S^{1})^{q}$, $q=D-p-1$%
. In the discussion below the length of the compact dimension $x^{l}$ will
be denoted by $L_{l}$ and $0\leqslant x^{l}\leqslant L_{l}$ for $l=p+2,...,D$%
. Hence, the subspace with the coordinates $(x^{2},x^{3},\ldots ,x^{D})$ has
spatial topology $R^{p}\times (S^{1})^{q}$. Introducing the coordinates%
\begin{equation}
t_{\mathrm{M}}=\xi \sinh \tau ,\quad x_{\mathrm{M}}^{1}=\xi \cosh \tau ,
\label{txM}
\end{equation}%
in the subspace $(\tau ,\xi )$, the line element (\ref{ds2}) is written in
the locally Minkowskian form
\begin{equation}
ds^{2}=\eta _{\mu \nu }dx_{\mathrm{M}}^{\mu }dx_{\mathrm{M}}^{\nu },
\label{ds2M}
\end{equation}%
where $\eta _{\mu \nu }=\mathrm{diag}(1,-1,\ldots ,-1)$ is the Minkowski
metric tensor and $x_{\mathrm{M}}^{0}=t_{\mathrm{M}}$, $x_{\mathrm{M}}^{\mu
}=x^{\mu }$ for $\mu =2,\ldots ,D$. As seen from (\ref{txM}), the
coordinates $x^{\mu }$ cover the right Rindler wedge $x_{\mathrm{M}}^{1}>|t_{%
\mathrm{M}}|$. The worldline for given $(\xi ,x^{2},\ldots ,x^{D})$
corresponds to an observer with constant proper acceleration $1/\xi $. The
proper time for that observer is expressed in terms of the dimensionless
time coordinate $\tau $ as $t=\xi \tau $.

Having specified the geometry we turn to the field content. We consider a
quantum charged scalar field $\varphi (x)$. Assuming the presence of an
external classical gauge field with the vector potential $A_{\mu }$, the
field equation reads%
\begin{equation}
\left( g^{\mu \nu }D_{\mu }D_{\nu }+m^{2}\right) \varphi =0,  \label{Feq}
\end{equation}%
where $D_{\mu }=\nabla _{\mu }+ieA_{\mu }$, the operator $\nabla _{\mu }$
stands for the standard covariant derivative in coordinates $x^{\mu }=\left(
\tau ,\xi ,\mathbf{x}_{p},\mathbf{x}_{q}\right) $ and $e$ is the charge of
the field quanta. Note that the background geometry is flat and the equation
(\ref{Feq}) is valid for general case of the curvature coupling parameter.
The spatial topology is nontrivial and for the theory to be defined in
addition to the field equation we need to specify the periodicity conditions
on the field operator along compact dimensions. In what follows generic
quasiperiodic conditions%
\begin{equation}
\varphi (t,\xi ,\mathbf{x}_{p},\mathbf{x}_{q}+L_{l}\mathbf{e}%
_{l})=e^{i\alpha _{l}}\varphi (t,\xi ,\mathbf{x}_{p},\mathbf{x}_{q}),
\label{PC}
\end{equation}%
will be imposed for $l=p+2,\ldots ,D$. In (\ref{PC}), $\mathbf{e}_{l}$ is
the unit vector along the compact dimension $x^{l}$ and the constant phases $%
\alpha _{l}$ can differ for different directions. The special cases of
untwisted and twisted fields, most frequently considered in the literature,
are realized by the choice $\alpha _{l}=0$ and $\alpha _{l}=\pi $,
respectively, for all values of $l=p+2,\ldots ,D$.

For an external gauge field $A_{\mu }$ we will take a simple configuration
with the only nonzero components $A_{\mu }=\mathrm{const}$ along compact
dimensions $x^{\mu }$, $\mu =p+2,\ldots ,D$. By making use of the gauge
transformation%
\begin{equation}
\varphi ^{\prime }(x)=e^{ie\chi }\varphi (x),\;A_{\mu }^{\prime }=A_{\mu
}-\partial _{\mu }\chi ,  \label{gtrans}
\end{equation}
with $\chi =A_{\mu }x^{\mu }$ we get $A_{\mu }^{\prime }=0$ and the vector
potential does no enter in the field equation for $\varphi ^{\prime }(x)$.
However, the gauge transformation modifies the periodicity conditions for
the new field as
\begin{equation}
\varphi ^{\prime }(t,\mathbf{x}_{p},\mathbf{x}_{q}+L_{l}\mathbf{e}_{l})=e^{i%
\tilde{\alpha}_{l}}\varphi ^{\prime }(t,\mathbf{x}_{p},\mathbf{x}_{q}),
\label{PCn}
\end{equation}%
with the new phases%
\begin{equation}
\tilde{\alpha}_{l}=\alpha _{l}+eA_{l}L_{l}.  \label{alftilde}
\end{equation}%
The physical observables depend on the phases $\alpha _{l}$ and on the
constant components $A_{l}$ through the gauge invariant combination (\ref%
{alftilde}). The effect of a constant gauge field is related to the
nontrivial spatial topology and is of the Aharonov-Bohm-type. The
consideration below will be presented in terms of the field $\varphi
^{\prime }(x)$. Omitting the prime, the corresponding field equation is
given by (\ref{Feq}) with $D_{\mu }=\nabla _{\mu }$.

The properties of the vacuum state in the problem under consideration are
completely determined by the two-point functions. For the evaluation of
those functions we will use the representation in the form of the summation
over complete set of mode functions. The mode functions realizing the FR
vacuum are specified by the energy $\omega $ and the momentum $\mathbf{k}$
and are given by the expression (for quantization of fields in Rindler
spacetime with trivial topology see, for example, \cite{Birr82,Cris08})
\begin{equation}
\varphi _{\omega ,\mathbf{k}}^{(\pm )}(x)=C_{\omega ,\mathbf{k}}e^{i\mathbf{k%
}\cdot \mathbf{x}\mp i\omega \tau }K_{i\omega }(\lambda \xi ),\;\lambda =%
\sqrt{k^{2}+m^{2}},  \label{fisigma}
\end{equation}%
where the upper and lower signs correspond to the positive and negative
energy modes, $\mathbf{x}=(\mathbf{x}_{p},\mathbf{x}_{q})$, $k=|\mathbf{k}|$%
, and $K_{\nu }(x)$ is the modified Bessel function \cite{Abra}. The
momentum $\mathbf{k}$ is decomposed as $\mathbf{k}=(\mathbf{k}_{p},\mathbf{k}%
_{q})$, with the components $\mathbf{k}_{p}=(k_{2},\ldots ,k_{p+1})$ and $%
\mathbf{k}_{q}=(k_{p+2},\ldots ,k_{D})$ corresponding to the uncopmact and
compact subspaces (with topologies $R^{p}$ and $(S^{1})^{q}$), respectively.
For the components along uncompact dimensions we have $-\infty
<k_{n}<+\infty $, $n=2,\ldots ,p+1$. The components along the compact
dimensions are quantized by the periodicity conditions (\ref{PCn}) with the
eigenvalues%
\begin{equation}
k_{l}=\left( 2\pi n_{l}+\tilde{\alpha}_{l}\right) /L_{l},  \label{kltild}
\end{equation}%
where $n_{l}=0,\pm 1,\pm 2,\ldots .,$ and $l=p+2,...,D$. Note that the
energy $\omega $ in (\ref{fisigma}) is dimensionless. The energy measured by
an observer with proper acceleration $1/\xi $ is expressed as $\omega /\xi $%
. The mode functions (\ref{fisigma}) are normalized by the condition
\begin{equation}
\int d{\mathbf{x}}\int_{0}^{\infty }\frac{d\xi }{\xi }\varphi _{\omega ,%
\mathbf{k}}^{(s)}\overset{\leftrightarrow }{\partial }_{\tau }\varphi
_{\omega ^{\prime },\mathbf{k}^{\prime }}^{(s^{\prime })\ast }=i\delta
_{ss^{\prime }}\delta \left( \omega -\omega ^{\prime }\right) \delta \left(
\mathbf{k}_{p}-\mathbf{k}_{p}^{\prime }\right) \delta _{\mathbf{k}_{q}%
\mathbf{k}_{q}^{\prime }},  \label{Norm}
\end{equation}%
where $\delta _{\mathbf{k}_{q}\mathbf{k}_{q}^{\prime }}=\delta
_{n_{p+2},n_{p+2}^{\prime }}....\delta _{n_{D},n_{D}^{\prime }}$. By using
the result (see, for example, \cite{Cand76})
\begin{equation}
\int_{0}^{\infty }\frac{dy}{y}\,K_{i\omega }(y)K_{i\omega ^{\prime }}(y)=%
\frac{\pi ^{2}\delta \left( \omega -\omega ^{\prime }\right) }{2\omega \sinh
(\omega \pi )},  \label{KKint}
\end{equation}%
for the coefficient in (\ref{fisigma}) we get
\begin{equation}
|C_{\omega ,\mathbf{k}}|^{2}=\frac{\sinh \left( \pi \omega \right) }{(2\pi
)^{p}\pi ^{2}V_{q}}.  \label{Cnorm}
\end{equation}%
Here and below, for the volume of the compact subspace we use the notation $%
V_{q}=L_{p+2}....L_{D}$. In the present paper we are interested in the VEV
of the current density (for quantum effects in models with toroidally
compact dimensions see \cite{Khan14} and references therein)%
\begin{equation}
j_{\mu }(x)=ie[\varphi ^{\dagger }(x)\partial _{\mu }\varphi (x)-(\partial
_{\mu }\varphi ^{\dagger }(x))\varphi (x)],  \label{jmu}
\end{equation}%
for the field $\varphi (x)$. It can be evaluated by using the two-point
function. As such a function in the next section we consider the Hadamard
function.

\section{Hadamard function}

\label{sec:Hadam}

The Hadamard function for a charged scalar field $\varphi (x)$ is defined as
the VEV
\begin{equation}
G(x,x^{\prime })=\left\langle 0\right\vert \varphi (x)\varphi ^{\dagger
}(x^{\prime })+\varphi ^{\dagger }(x^{\prime })\varphi (x)\left\vert
0\right\rangle ,  \label{G1}
\end{equation}%
with $\left\vert 0\right\rangle $ being the vacuum state (the FR vacuum in
our consideration). Having the mode functions, the Hadamard function for the
FR vacuum can be written in the form of the mode sum%
\begin{equation}
G(x,x^{\prime })=\int d\mathbf{k}_{p}\int_{0}^{\infty }d\omega \sum_{\mathbf{%
n}_{q}}\sum_{s=\pm }\varphi _{\mathbf{k}}^{(s)}(x)\varphi _{\mathbf{k}%
}^{(s)\ast }(x^{\prime }),  \label{G01}
\end{equation}%
where $\mathbf{n}_{q}=(n_{p+2},n_{p+3},\ldots ,n_{D})$ and%
\begin{equation}
\sum_{\mathbf{n}_{q}}=\sum_{n_{p+2}=-\infty }^{+\infty }\cdots
\sum_{n_{D}=-\infty }^{+\infty }.  \label{Sumnq}
\end{equation}%
Substituting the mode functions from (\ref{fisigma}) we get%
\begin{equation}
G(x,x^{\prime })=\frac{2^{1-p}}{\pi ^{p+2}V_{q}}\int d{\mathbf{k}}%
_{p}\,\sum_{\mathbf{n}_{q}}e^{i{\mathbf{k}}\cdot \Delta {\mathbf{x}}%
}\int_{0}^{\infty }d\omega \sinh (\pi \omega )\cos \left( \omega \Delta \tau
\right) K_{i\omega }(\lambda _{\mathbf{k}}\xi )K_{i\omega }(\lambda _{%
\mathbf{k}}\xi ^{\prime }),  \label{emtRindler}
\end{equation}%
where $\Delta {\mathbf{x=x}}-{\mathbf{x}^{\prime }}$, $\Delta \tau =\tau
-\tau ^{\prime }$ and%
\begin{equation}
\lambda _{\mathbf{k}}=\sqrt{\mathbf{k}_{p}^{2}+\mathbf{k}_{q}^{2}+m^{2}},\;%
\mathbf{k}_{q}^{2}=\sum_{l=p+2}^{D}\left( \frac{2\pi n_{l}+\tilde{\alpha}_{l}%
}{L_{l}}\right) ^{2}.  \label{lamk}
\end{equation}

In order to compare the effects of compactification on the FR and Minkowski
vacua let us consider the Hadamard function for the Minkowski vacuum $%
\left\vert 0\right\rangle _{\mathrm{M}}$ in the locally Minkowski spacetime
with the line element (\ref{ds2M}) and the spatial topology $R^{p+1}\times
(S^{1})^{q}$ (for a recent discussion of relations between the Minkowski and
Rindler propagators see \cite{Raje20} and references therein). In Appendix %
\ref{sec:App2} we show that the Minkowskian Hadamard function $G_{\mathrm{M}%
}(x,x^{\prime })$ is presented in the form
\begin{equation}
G_{\mathrm{M}}(x,x^{\prime })=\frac{2^{1-p}}{\pi ^{p+2}V_{q}}\int d\mathbf{k}%
_{p}\sum_{\mathbf{n}_{q}}e^{i\mathbf{k}\cdot \Delta {\mathbf{x}}%
}\int_{0}^{\infty }d\omega \,\cosh \left( \pi \omega \right) \cos \left(
\omega \Delta \tau \right) K_{i\omega }(\lambda _{\mathbf{k}}\xi )K_{i\omega
}(\lambda _{\mathbf{k}}\xi ^{\prime }).  \label{Gmin2}
\end{equation}%
This form is convenient for the investigation of the difference between the
VEVs in the FR and Minkowski vacua.

By using the representations (\ref{emtRindler}) and (\ref{Gmin2}), the
difference of the Hadamard functions is expressed as%
\begin{eqnarray}
G(x,x^{\prime })-G_{\mathrm{M}}(x,x^{\prime }) &=&-\frac{2^{1-p}}{\pi
^{p+2}V_{q}}\int d{\mathbf{k}}_{p}\,\sum_{\mathbf{n}_{q}}e^{i{\mathbf{k}}%
\cdot \Delta {\mathbf{x}}}\int_{0}^{\infty }d\omega \,e^{-\pi \omega }
\notag \\
&&\times \cos \left( \omega \Delta \tau \right) K_{i\omega }(\lambda _{%
\mathbf{k}}\xi )K_{i\omega }(\lambda _{\mathbf{k}}\xi ^{\prime }).
\label{Gdif}
\end{eqnarray}%
For the further transformation of (\ref{Gdif}) we employ the integral
representation \cite{Wats66} (in \cite{Wats66} there is a missprint, instead
of $\left( x^{2}+X^{2}\right) /u$ should be $-\left( x^{2}+X^{2}\right) /2u$)%
\begin{equation}
K_{\nu }(X)K_{\nu }(x)=\frac{1}{2}\int_{0}^{\infty }dT\int_{0}^{\infty }%
\frac{du}{u}\cosh \left( \nu T\right) e^{-(xX/u)\cosh T}\exp \left[ -\left(
\frac{u}{2}+\frac{x^{2}+X^{2}}{2u}\right) \right] ,  \label{IntRepKK}
\end{equation}%
for the product of the modified Bessel functions. Substituting (\ref%
{IntRepKK}) in (\ref{Gdif}), we first integrate over $\omega $ and then over
$u$ and ${\mathbf{k}}_{p}$. This leads to the final result%
\begin{eqnarray}
G(x,x^{\prime }) &=&G_{\mathrm{M}}(x,x^{\prime })-\frac{2^{-p/2}}{\pi
^{p/2+1}V_{q}}\sum_{\mathbf{n}_{q}}\omega _{\mathbf{n}_{q}}^{p}e^{i{\mathbf{k%
}}_{q}\cdot ({\mathbf{x}}_{q}-{\mathbf{x}_{q}^{\prime }})}\int_{0}^{\infty
}dT\sum_{s=\pm 1}\frac{1}{\left( T-s\Delta \tau \right) ^{2}+\pi ^{2}}
\notag \\
&&\times f_{p/2}\left( \omega _{\mathbf{n}_{q}}\sqrt{|{\mathbf{x}}_{p}-{%
\mathbf{x}_{p}^{\prime }}|^{2}+\xi ^{2}+\xi ^{\prime 2}+2\xi \xi ^{\prime
}\cosh T}\right) ,  \label{Gdif2}
\end{eqnarray}%
where we have introduced the notations $f_{\nu }(x)=K_{\nu }(x)/x^{\nu }$
and $\omega _{\mathbf{n}_{q}}=\sqrt{\mathbf{k}_{q}^{2}+m^{2}}$. Note that
the divergences for $G(x,x^{\prime })$ and $G_{\mathrm{M}}(x,x^{\prime })$
in the coincidence limit $x^{\prime }\rightarrow x$ are the same and the
last term in (\ref{Gdif2}) is finite in that limit.

\section{Current density}

\label{sec:Current}

\subsection{General expression}

Having the Hadamard function we can evaluate the expectation value for the
current density (\ref{jmu}) in the FR vacuum by using the formula%
\begin{equation}
\left\langle 0\right\vert j_{\mu }(x)\left\vert 0\right\rangle \equiv
\left\langle j_{\mu }\right\rangle =\frac{i}{2}e\lim_{x^{\prime }\rightarrow
x}(\partial _{\mu }-\partial _{\mu }^{\prime })G(x,x^{\prime }).  \label{jW}
\end{equation}%
First we can see that the VEVs for the charge density and for the components
along the uncompact directions vanish, $\left\langle j_{\mu }\right\rangle
=0 $, $\mu =0,1,\ldots ,p+1$. By making use of (\ref{jW}) and the expression
(\ref{Gdif2}) of the Hadamard function, for the contravariant component of
the current density along the $l$-th compact dimension we get:%
\begin{equation}
\langle j^{l}\rangle =\langle j^{l}\rangle _{\mathrm{M}}-\frac{2^{-p/2}e}{%
\pi ^{p/2+1}V_{q}}\sum_{\mathbf{n}_{q}}k_{l}\omega _{\mathbf{n}%
_{q}}^{p}\int_{0}^{\infty }dx\,\frac{f_{p/2}\left( 2\xi \omega _{\mathbf{n}%
_{q}}\cosh x\right) }{x^{2}+\pi ^{2}/4},  \label{jr}
\end{equation}%
where $k_{l}$ is given by (\ref{kltild}) and $\langle j^{l}\rangle _{\mathrm{%
M}}$ is the current density for the Minkowski vacuum. The latter has been
investigated in \cite{Beze13}. The corresponding formula is obtained from
\cite{Beze13} by the replacement $p\rightarrow p+1$:%
\begin{equation}
\langle j^{l}\rangle _{\mathrm{M}}=\frac{4eL_{l}^{2}V_{q}^{-1}}{(2\pi
)^{\left( p+3\right) /2}}\sum_{n_{l}=1}^{\infty }n_{l}\sin (n_{l}\tilde{%
\alpha}_{l})\sum_{\mathbf{n}_{q-1}^{l}}\omega _{\mathbf{n}_{q-1}}^{p+3}f_{%
\frac{p+3}{2}}(n_{l}L_{l}\omega _{\mathbf{n}_{q-1}}),  \label{jrM}
\end{equation}%
where $\mathbf{n}_{q-1}^{l}=(n_{p+2},\ldots ,n_{l-1},,n_{l+1},\ldots ,n_{D})$
and $\omega _{\mathbf{n}_{q-1}}=\sqrt{\mathbf{k}_{q-1}^{2}+m^{2}}$ with $%
\mathbf{k}_{q-1}^{2}=\mathbf{k}_{q}^{2}-k_{l}^{2}$. An equivalent expression
is given by \cite{Beze13}%
\begin{equation}
\langle j^{l}\rangle _{\mathrm{M}}=\frac{2eL_{l}m^{D+1}}{(2\pi )^{(D+1)/2}}%
\sum_{\mathbf{n}_{q}}n_{l}\sin \left( \mathbf{n}_{q}\cdot \boldsymbol{\tilde{%
\alpha}}\right) f_{\frac{D+1}{2}}\left( m\sqrt{\sum%
\nolimits_{i=p+2}^{D}n_{i}^{2}L_{i}^{2}}\right) ,  \label{jrM2}
\end{equation}%
with $\mathbf{n}_{q}\cdot \boldsymbol{\tilde{\alpha}}=\sum%
\nolimits_{i=p+2}^{D}n_{i}\tilde{\alpha}_{i}$. Note that the Minkowskian
part is homogeneous. Both the contributions in (\ref{jr}) are odd periodic
functions of $\tilde{\alpha}_{l}$ with the period $2\pi $ and even periodic
functions of $\tilde{\alpha}_{i}$, $i\neq l$,with the same period. By making
use of a gauge transformation similar to (\ref{gtrans}) with $\chi \left(
x\right) =-\sum_{i=p+2}^{D}\alpha _{i}x^{i}/eL_{i}$ we can pass to a new
gauge with the fields $(\tilde{\varphi}(x),\tilde{A}_{\mu })$, where $\tilde{%
A}_{i}=\tilde{\alpha}_{i}/eL_{i}$, $i=p+2,\ldots ,D$, and the function $%
\tilde{\varphi}(x)$ is periodic along compact dimensions. In this gauge, the
formal magnetic flux enclosed by the $i$th compact dimension is given as $%
\tilde{\Phi}_{i}=-\tilde{A}_{i}L_{i}$. The parameter $\tilde{\alpha}_{i}$ is
expressed as $\tilde{\alpha}_{i}=-2\pi \tilde{\Phi}_{i}/\Phi _{0}$, where $%
\Phi _{0}=2\pi /e$ ($2\pi \hbar c/e$ in standard units) is the flux quantum.
As seen, the periodicity with respect to $\tilde{\alpha}_{i}$ is translated
to a periodicity with respect to the magnetic flux $\tilde{\Phi}_{i}$ with
the period of flux quantum.

An alternative expression for the current density $\langle j^{l}\rangle $ in
the FR vacuum is obtained by using the formula \cite{Beze13}%
\begin{equation}
\sum_{\mathbf{n}}\cos (\mathbf{n}\cdot \boldsymbol{\beta })f_{\nu }(c\sqrt{%
b^{2}+\sum\nolimits_{i=1}^{r}a_{i}^{2}n_{i}^{2}})=\frac{(2\pi )^{r/2}}{%
a_{1}\cdots a_{r}c^{2\nu }}\sum_{\mathbf{n}}w_{\mathbf{n}}^{2\nu -r}f_{\nu -%
\frac{r}{2}}(bw_{\mathbf{n}}),  \label{Rel4}
\end{equation}%
where $\mathbf{n}=(n_{1},\ldots ,n_{r})$, $\boldsymbol{\beta }=(\beta
_{1},\ldots ,\beta _{r})$, and $w_{\mathbf{n}}^{2}=\sum\nolimits_{i=1}^{r}(2%
\pi n_{i}+\beta _{i})^{2}/a_{i}^{2}+c^{2}$. By taking the derivative with
respect to $\beta _{l}$ from here we find
\begin{equation}
\sum_{\mathbf{n}}\frac{2\pi n_{l}+\beta _{l}}{a_{l}^{2}}w_{\mathbf{n}}^{2\nu
-r-2}f_{\nu -\frac{r}{2}-1}(bw_{\mathbf{n}})=\frac{a_{1}\cdots a_{r}c^{2\nu }%
}{(2\pi )^{r/2}}\sum_{\mathbf{n}}n_{l}\sin (\mathbf{n}\cdot \boldsymbol{%
\beta })f_{\nu }(c\sqrt{b^{2}+\sum\nolimits_{i=1}^{r}a_{i}^{2}n_{i}^{2}}).
\label{Rel5}
\end{equation}%
Choosing $r=D-p-1$, $c=m$, $\nu =(D+1)/2$, $\beta _{i}=\tilde{\alpha}%
_{p+i+1} $ and $a_{i}=L_{p+i+1}$, one gets the formula
\begin{equation}
\sum_{\mathbf{n}_{q}}k_{l}\omega _{\mathbf{n}_{q}}^{p}f_{\frac{p}{2}%
}(b\omega _{\mathbf{n}_{q}})=\frac{V_{q}L_{l}m^{D+1}}{(2\pi )^{(D-p-1)/2}}%
\sum_{\mathbf{n}_{q}}n_{l}\sin (\mathbf{n}_{q}\cdot \boldsymbol{\tilde{\alpha%
}})f_{\frac{D+1}{2}}(m\sqrt{b^{2}+\sum\nolimits_{i=p+2}^{D}L_{i}^{2}n_{i}^{2}%
}).  \label{Rel6}
\end{equation}%
By using this relation with $b=2\xi \cosh x$ in (\ref{jr}), the following
formula is obtained%
\begin{eqnarray}
\langle j^{l}\rangle &=&\langle j^{l}\rangle _{\mathrm{M}}-\frac{%
2em^{D+1}L_{l}}{(2\pi )^{(D+1)/2}}\sum_{\mathbf{n}_{q}}n_{l}\sin \left(
\mathbf{n}_{q}\cdot \boldsymbol{\tilde{\alpha}}\right)  \notag \\
&&\times \int_{0}^{\infty }\,\frac{dx}{x^{2}+\pi ^{2}/4}f_{\frac{D+1}{2}}(m%
\sqrt{4\xi ^{2}\cosh ^{2}x+\sum\nolimits_{i=p+2}^{D}L_{i}^{2}n_{i}^{2}}).
\label{jr2}
\end{eqnarray}%
Note that in this formula we can make the replacement
\begin{equation}
\sum_{\mathbf{n}_{q}}n_{l}\sin \left( \mathbf{n}_{q}\cdot \boldsymbol{\tilde{%
\alpha}}\right) \rightarrow 2\sum_{n_{l}=1}^{\infty }n_{l}\sin (n_{l}\tilde{%
\alpha}_{l})\sum_{\mathbf{n}_{q-1}^{l}}\cos \left( \sum\nolimits_{i=p+2,\neq
l}^{D}n_{i}\tilde{\alpha}_{i}\right) .  \label{Repl}
\end{equation}

The general expression (\ref{jr2}) is further simplified for a massless
field. By taking into account that%
\begin{equation}
f_{\nu }(x)\approx 2^{\nu -1}\Gamma (\nu )x^{-2\nu },\;x\ll 1,
\label{fsmall}
\end{equation}%
we get%
\begin{eqnarray}
\langle j^{l}\rangle &=&\langle j^{l}\rangle _{\mathrm{M}}-\frac{e\Gamma
((D+1)/2)}{\pi ^{(D+1)/2}}L_{l}\sum_{\mathbf{n}_{q}}n_{l}\sin \left( \mathbf{%
n}_{q}\cdot \boldsymbol{\tilde{\alpha}}\right)  \notag \\
&&\times \int_{0}^{\infty }\,\frac{dx}{x^{2}+\pi ^{2}/4}(4\xi ^{2}\cosh
^{2}x+\sum\nolimits_{i=p+2}^{D}L_{i}^{2}n_{i}^{2})^{-\frac{D+1}{2}}.
\label{jrm0}
\end{eqnarray}%
We can also make the replacement (\ref{Repl}). The corresponding formula for
the Minkowskian part is obtained from (\ref{jrM2}):%
\begin{equation*}
\langle j^{l}\rangle _{\mathrm{M}}=\frac{e\Gamma ((D+1)/2)}{\pi ^{(D+1)/2}}%
L_{l}\sum_{\mathbf{n}_{q}}n_{l}\sin \left( \mathbf{n}_{q}\cdot \boldsymbol{%
\tilde{\alpha}}\right) \left(
\sum\nolimits_{i=p+2}^{D}n_{i}^{2}L_{i}^{2}\right) ^{-\frac{D+1}{2}}.
\end{equation*}

The numerical results below will be given for the geometry with a single
compact dimension $x^{D}$ of the length $L$. In this special case one has $%
q=1$, $p=D-2$ and the general formulas are simplified to
\begin{equation}
\langle j^{D}\rangle =\langle j^{D}\rangle _{\mathrm{M}}-\frac{2^{1-D/2}e}{%
\pi ^{D/2}L}\sum_{n=-\infty }^{+\infty }k_{D}\omega
_{D}^{D-2}\int_{0}^{\infty }dx\,\frac{f_{D/2-1}\left( 2\xi \omega _{D}\cosh
x\right) }{x^{2}+\pi ^{2}/4},  \label{jrq1}
\end{equation}%
with $k_{D}=\left( 2\pi n+\tilde{\alpha}_{D}\right) /L$ and $\omega _{D}=%
\sqrt{k_{D}^{2}+m^{2}}$. The Minkowskian part takes the form%
\begin{equation}
\langle j^{D}\rangle _{\mathrm{M}}=\frac{4em^{D+1}L}{(2\pi )^{(D+1)/2}}%
\sum_{n=1}^{\infty }n\sin (n\tilde{\alpha}_{D})f_{\frac{D+1}{2}}\left(
nmL\right) .  \label{jrq1M}
\end{equation}%
Note that in this special case the representations (\ref{jrM}) and (\ref%
{jrM2}) coincide. An equivalent representation is obtained from (\ref{jr2}):%
\begin{equation}
\langle j^{D}\rangle =\langle j^{D}\rangle _{\mathrm{M}}-\frac{4em^{D+1}L}{%
(2\pi )^{(D+1)/2}}\sum_{n=1}^{\infty }n\sin \left( n\tilde{\alpha}%
_{D}\right) \int_{0}^{\infty }\,dx\,\frac{f_{(D+1)/2}(m\sqrt{4\xi ^{2}\cosh
^{2}x+n^{2}L^{2}})}{x^{2}+\pi ^{2}/4}.  \label{jrq1b}
\end{equation}%
For a massless field this is reduced to
\begin{equation}
\langle j^{D}\rangle =\langle j^{D}\rangle _{\mathrm{M}}-2eL\frac{\Gamma
((D+1)/2)}{\pi ^{(D+1)/2}}\sum_{n=1}^{\infty }n\sin \left( n\tilde{\alpha}%
_{D}\right) \int_{0}^{\infty }\,dx\frac{(4\xi ^{2}\cosh ^{2}x+n^{2}L^{2})^{-%
\frac{D+1}{2}}}{x^{2}+\pi ^{2}/4},  \label{jrq1m0}
\end{equation}%
with the Minkowskian current density%
\begin{equation}
\langle j^{D}\rangle _{\mathrm{M}}=\frac{2e\Gamma ((D+1)/2)}{\pi
^{(D+1)/2}L^{D}}\sum_{n=1}^{\infty }\frac{\sin (n\tilde{\alpha}_{D})}{n^{D}}.
\label{jDMm0}
\end{equation}

\subsection{Asymptotic analysis and numerical results}

Let us consider the behavior of the current density in the asymptotic
regions of the coordinate $\xi $. For points near the Rindler horizon one
has $\xi \ll L_{i}$. In this limit it is convenient to use the
representation (\ref{jr2}). Directly putting $\xi =0$, the integral over $x$
gives 1 and we can see that the leading term in the expansion over $\xi $ of
the last term in (\ref{jr2}) coincides with the current density $\langle
j^{l}\rangle _{\mathrm{M}}$. From here we conclude that the current density $%
\langle j^{l}\rangle $ vanishes on the Rindler horizon. In the opposite
limit $\xi \gg L_{i}$ it is more convenient to use the representation (\ref%
{jr}). By using the asymptotic expression for the modified Bessel function
for large arguments \cite{Abra}, we see that the dominant contribution to
the series over $\mathbf{n}_{q}$ comes from the term with the smallest value
of $\omega _{\mathbf{n}_{q}}$ that will be denoted here by $\omega _{0}$.
Assuming that $|\tilde{\alpha}_{i}|\leqslant \pi $, one has%
\begin{equation}
\omega _{0}=\sqrt{\sum\nolimits_{i=p+2}^{D}\tilde{\alpha}%
_{i}^{2}/L_{i}^{2}+m^{2}}.  \label{om0}
\end{equation}%
In addition, the main contribution to the integral over $x$ comes from the
integration range near the lower limit. In this way we can see that, to the
leading order,
\begin{equation}
\langle j^{l}\rangle \approx \langle j^{l}\rangle _{\mathrm{M}}-\frac{e%
\tilde{\alpha}_{l}\omega _{0}^{p/2-1}e^{-2\xi \omega _{0}}}{2^{p}\pi
^{p/2+2}V_{q}L_{l}\xi ^{p/2+1}},  \label{jrlarge}
\end{equation}%
and the difference of the current densities in the Minkowski and FR vacua is
exponentially suppressed.

Now we turn to the asymptotics with respect to the length $L_{l}$ of the
compact dimension. Under the condition $L_{l}\ll \xi $, the behavior of the
current density is described by (\ref{jrlarge}). If in addition one has $%
L_{l}\ll 1/m$, we can substitute $\omega _{0}\approx |\tilde{\alpha}%
_{l}|/L_{l}$. The asymptotic for the Minkowskian part $\langle j^{l}\rangle
_{\mathrm{M}}$ has been discussed in \cite{Beze13}. The leading term
coincides with the current density for a massless scalar field in the space
with topology $R^{D-1}\times S^{1}$. It is obtained from (\ref{jrq1M}) in
the limit $m\rightarrow 0$ and making the replacements $\tilde{\alpha}%
_{D}\rightarrow \tilde{\alpha}_{l}$, $L\rightarrow L_{l}$. This shows that
the Minkowskian part behaves as $1/L_{l}^{D}$ and it dominates in the
asymptotic region under consideration.

For large values of $L_{l}$, compared with other length scales, and for $m=0$%
, $\tilde{\alpha}_{i}=0$, $i\neq l$, the Minkowskian part behaves as%
\begin{equation}
\langle j^{l}\rangle _{\mathrm{M}}\approx \frac{2e\Gamma ((p+3)/2)}{\pi
^{(p+3)/2}L_{l}^{p+1}V_{q}}\sum_{n_{l}=1}^{\infty }\frac{\sin (n_{l}\tilde{%
\alpha}_{l})}{n_{l}^{p+2}}.  \label{jMlargeL}
\end{equation}%
The behavior of $\langle j^{l}\rangle _{\mathrm{M}}$ is essentially
different for $\omega _{0l}=\sqrt{\sum\nolimits_{i=p+1,\neq l}^{D}\tilde{%
\alpha}_{i}^{2}/L_{i}^{2}+m^{2}}\neq 0$. In this case the suppression of the
current density $\langle j^{l}\rangle _{\mathrm{M}}$ is exponential:%
\begin{equation}
\langle j^{l}\rangle _{\mathrm{M}}\approx \frac{2e\omega _{0l}^{p/2+1}\sin
\tilde{\alpha}_{l}}{(2\pi )^{p/2+1}V_{q}^{1}L_{l}^{p/2}}e^{-L_{l}\omega
_{0l}}.  \label{jMlargeL2}
\end{equation}%
Considering the part induced by acceleration, for large values of $L_{l}$,
in the leading order, we can ignore the term $4\xi ^{2}\cosh ^{2}x$ under
the sign of square root in (\ref{jr2}). The integral over $x$ gives 1 and
the leading term in the expansion of the difference $\langle j^{l}\rangle
-\langle j^{l}\rangle _{\mathrm{M}}$ coincides with that for $-\langle
j^{l}\rangle _{\mathrm{M}}$ (see (\ref{jrM2})). Hence, the decay of the
current density $\langle j^{l}\rangle $ for large $L_{l}$ is stronger than
that for $\langle j^{l}\rangle _{\mathrm{M}}$. For large values of $L_{i}$, $%
i\neq l$, the main contribution in (\ref{jr2}) comes from the term with $%
n_{i}=0$ and the leading term in the expansion for $\langle j^{l}\rangle $
coincides with the current density in the model where the dimension $x^{i}$
is decompactified.

\begin{figure}[tbph]
\begin{center}
\epsfig{figure=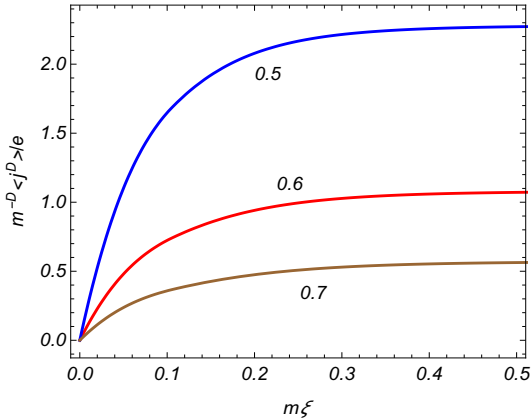,width=7.cm,height=5.5cm}
\end{center}
\caption{The current density along the compact dimension in the $D=4$ model
as a function of $m \protect\xi $. The graphs are plotted for $\tilde{%
\protect\alpha}_{D}/2\protect\pi =0.2$ and the numbers near the curves
correspond to the values of $mL$.}
\label{fig1}
\end{figure}

In figures below the graphs are plotted for the model $D=4$ with a single
compact dimension of the length $L$. Figure \ref{fig1} presents the
dependence of the current density on $m\xi $ for different values of $mL$
(the numbers near the curves) and for $\tilde{\alpha}_{D}/2\pi =0.2$. In the
limit $m\xi \rightarrow \infty $ the current density tends to $\langle
j^{D}\rangle _{\mathrm{M}}$. As it has been shown above by the asymptotic
analysis, the current density vanishes on the Rindler horizon $\xi =0$.

\begin{figure}[tbph]
\begin{center}
\begin{tabular}{cc}
\epsfig{figure=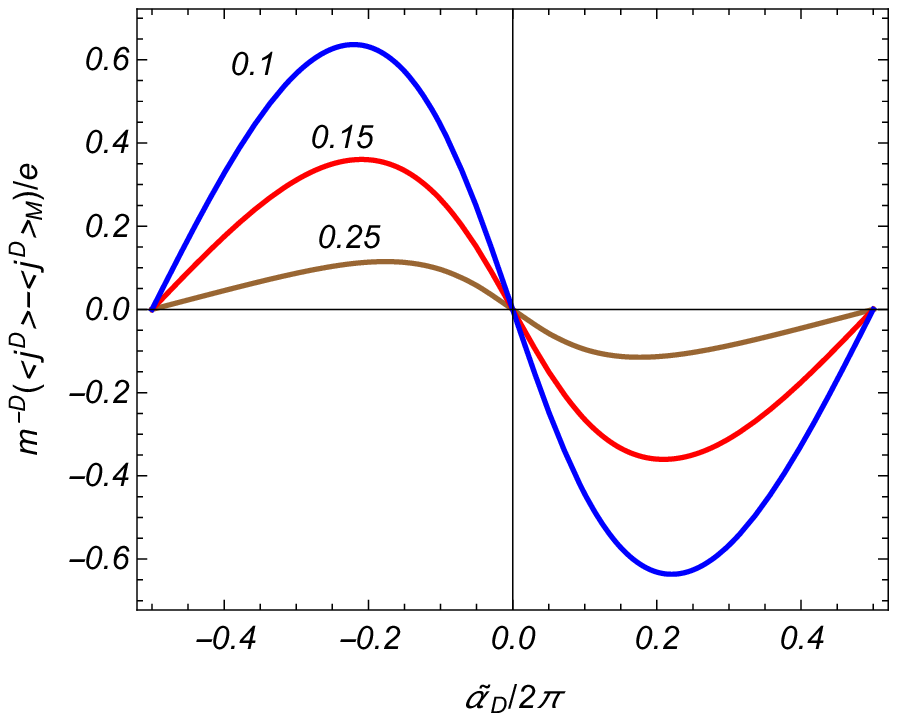,width=7.cm,height=5.5cm} & \quad %
\epsfig{figure=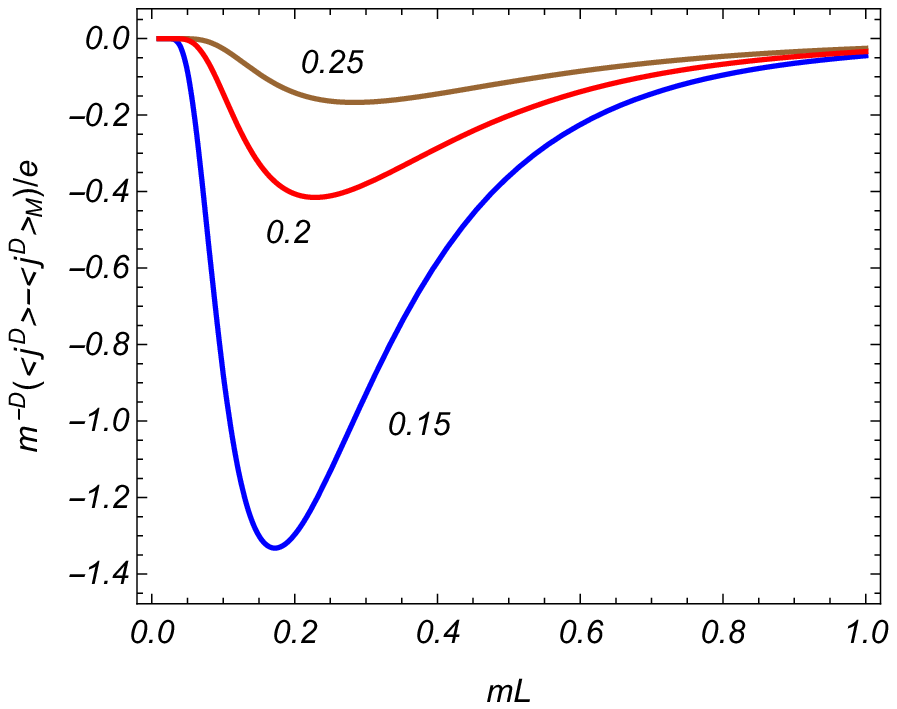,width=7.cm,height=5.5cm}%
\end{tabular}%
\end{center}
\caption{The dependence of the difference of the current densities in the FR
and Minkowski vacua on the parameter $\tilde{\protect\alpha}_{D}/2\protect%
\pi $ (left panel) and on the length of the compact dimension (right panel).
The left and right panels are plotted for $mL=0.5$ and $\tilde{\protect\alpha%
}_{D}/2\protect\pi =0.2$, respectively, and the numbers near the curves
correspond to the values of $m\protect\xi $. }
\label{fig2}
\end{figure}

In Figure \ref{fig2} we have plotted the difference in the current densities
for the FR and Minkowski vacua (in units of $em^{D}$) versus the parameter $%
\tilde{\alpha}_{D}/2\pi $ (left panel) and the length of the compact
dimension (right panel) in the model with $D=4$ for a single compact
dimension. The numbers near the curves are the values for $m\xi $. On the
left panel, for the length of the compact dimension we have taken the value
corresponding to $mL=0.5$. The current density is a periodic function of $%
\tilde{\alpha}_{D}/2\pi $ with the period 1 and the graphs are plotted for a
single period. We recall that $\tilde{\alpha}_{D}/2\pi $ is expressed in
terms of the magnetic flux threading the compact dimension in units of the
flux quantum. The graphs on the right panel are plotted for $\tilde{\alpha}%
_{D}/2\pi =0.2$. As it has been shown by asymptotic analysis, for $L\ll \xi $
the difference $\langle j^{D}\rangle -\langle j^{D}\rangle _{\mathrm{M}}$ is
suppressed by the factor $e^{-2|\tilde{\alpha}_{D}|\xi /L}$ and that is
confirmed by the numerical data on the right panel.

\subsection{Near-horizon and large-distance vacuum currents around
cylindrical black holes}

The cylindrical black holes are axially symmetric solutions of the Einstein
equations with negative cosmological constant (see, for example, \cite%
{Lemo95}-\cite{Gaet17} and references therein). They have also been studied
in the context of cosmic strings, supergravity and low energy string
theories. The line element for the exterior geometry of a non-rotating and
uncharged cylindrical black hole in $(D+1)$-dimensional spacetime has the
form (the generalization for rotating and charged black holes can be found
in \cite{Awad03})%
\begin{equation}
ds_{\mathrm{bh}}^{2}=f(r)dt^{2}-\frac{dr^{2}}{f(r)}-r^{2}\sum_{i=1}^{p}%
\left( dy^{i}\right) ^{2}-r^{2}\sum_{i=1}^{q}\left( d\phi _{i}\right) ^{2},
\label{ds2bh}
\end{equation}%
where $p+q+1=D$, $0\leq \phi _{i}<2\pi $, $-\infty <y^{i}<+\infty $, and
\begin{equation}
f(r)=\frac{r^{2}}{a^{2}}-\frac{M}{r^{D-2}}.  \label{fr}
\end{equation}%
The parameters $a$ and $M$ are expressed in terms of negative cosmological
constant $\Lambda $ and mass $\mathcal{M}$ per unit volume of the subspace $%
(y_{1},\ldots ,y_{p})$ as
\begin{equation}
a=\sqrt{D\frac{1-D}{2\Lambda }},\;M=\frac{16\pi G_{D+1}}{D-1}\mathcal{M},
\label{aM}
\end{equation}%
where $G_{D+1}$ is the Newton gravitational constant in $(D+1)$-dimensional
spacetime. The gravitational field is characterized by two length scales: $a$
and $r_{G}=M^{1/(D-2)}$. For the event horizon one has $r=r_{H}$ with $%
r_{H}=\left( a^{2}M\right) ^{1/D}=r_{G}(a/r_{G})^{2/D}$. In the special case
of a $D=3$ cylindrical black hole one has $p=q=1$. For $(2+1)$-dimensional
black holes $p=0$ and $q=1$. Some approximate and numerical results for the
VEVs\ of the field squared and energy-momentum tensor for a massless real
scalar field in the geometry of 4-dimensional cylindrical black hole were
presented in \cite{Bene98,Bene99}.

In Appendix \ref{sec:App1} we show that in the near-horizon limit the line
element (\ref{ds2bh}) takes a locally Rindler form (\ref{ds2}) with the
lengths of the compact dimensions determined by the radius of the horizon $%
r_{H}$: $L_{l}=2\pi r_{H}$ for $l=p+2,\ldots ,D$. From here we conclude that
near the horizon the current density along the direction $\phi _{i}$ is
expressed by the formulas (\ref{jr}) or (\ref{jr2}) with $l=i+p+1$ and
\begin{equation}
\xi =\frac{2a}{\sqrt{D}}\sqrt{r/r_{H}-1}.  \label{ksibh}
\end{equation}%
Note that these formulas give the physical components. The contravariant
components in the coordinate system $(t,r,y_{1},\ldots ,y_{p},\phi
_{1},\ldots ,\phi _{q})$ contain an additional factor $1/r$ that should be
replaced by $1/r_{H}$ in the near-horizon limit. From the analysis presented
in Appendix \ref{sec:App1} it follows that the expressions (\ref{jr}) or (%
\ref{jr2}) approximate the near-horizon currents for cylindrical black holes
under the condition $\left( \xi /a\right) ^{2}\ll 1$. For the ratio $\xi
/L_{l}$ with $L_{l}=2\pi r_{H}$ one gets $\xi /L_{l}\ll \left(
a/r_{G}\right) ^{1-2/D}/(2\pi )$. If the length scales $a$ and $r_{G}$ have
the same order of magnitude we have $\xi /L_{l}\ll 1$ and the corresponding
asymptotics of the formulas (\ref{jr}) or (\ref{jr2}) can be used to
estimate the current density. In the case $a\gg r_{G}$ the ratio $\xi /L_{l}$
can be of the order of 1 and the exact expressions should be used.

Now let us consider the asymptotic geometry at large distances from the
horizon, $r\gg r_{H}$. Expanding the metric tensor and introducing a new
coordinate $z=a^{2}/r$, the line element is expressed as
\begin{equation}
ds_{\mathrm{bh}}^{2}\approx \frac{a^{2}}{z^{2}}\left[ dt^{2}-dz^{2}-a^{2}%
\sum_{i=1}^{p}\left( dy^{i}\right) ^{2}-a^{2}\sum_{i=1}^{q}\left( d\phi
_{i}\right) ^{2}\right] .  \label{ds2bh1}
\end{equation}%
With the new rescaled coordinates $x^{i+1}=ay^{i}$, $i=1,\ldots ,p$, and $%
x^{p+i+1}=a\phi _{i}$, $i=1,\ldots ,q$, it attains the form%
\begin{equation}
ds_{\mathrm{bh}}^{2}\approx \frac{a^{2}}{z^{2}}\left[ dt^{2}-dz^{2}-%
\sum_{i=2}^{D}\left( dx^{i}\right) ^{2}\right] ,  \label{ds2bh2}
\end{equation}%
with $-\infty <x^{i}<+\infty $ for $i=2,\ldots ,p+1$, and $0\leq x^{i}\leq
2\pi a$ for $i=p+2,\ldots ,D$. The expression in the right-hand side of (\ref%
{ds2bh2}) describes a locally AdS spacetime a part of the coordinates for
which are compactified to a torus. The values $z=0$ and $z=\infty $ for the $%
z$-coordinate correspond to the AdS boundary and horizon, respectively. From
the condition $r\gg r_{H}$ it follows that the line element in (\ref{ds2bh2}%
) describes the large-distance asymptotic for cylindrical black holes under
the condition $z/a\ll (a/r_{G})^{1-2/D}$. For $a$ and $r_{G}$ having the
same order of magnitude this corresponds to points near the AdS boundary.

The VEV\ of the current density for a scalar field in the geometry described
by the right-hand side of (\ref{ds2bh2}) with general values $L_{l}$ of the
lengths of compact dimensions has been investigated in \cite{Beze15}. In the
special case under consideration with $L_{l}=2\pi a$ it is given by the
expression
\begin{equation}
\langle j^{l}\rangle \approx \frac{2ea^{-D}}{(2\pi )^{(D-1)/2}}\sum_{\mathbf{%
n}_{q}}n_{l}\sin \left( \mathbf{n}_{q}\cdot \boldsymbol{\tilde{\alpha}}%
\right) q_{\nu -1/2}^{(D+1)/2}(1+2\left( \pi r/a\right)
^{2}\sum_{i=p+2}^{D}n_{i}^{2}),  \label{jlAdS}
\end{equation}%
where $\nu =\sqrt{D^{2}/4-D(D+1)\zeta +m^{2}a^{2}}$ and $q_{\alpha }^{\mu
}(x)=(x^{2}-1)^{-\mu /2}e^{-i\pi \mu }Q_{\alpha }^{\mu }(x)$. Here, $\zeta $
is the curvature coupling parameter and $Q_{\alpha }^{\mu }(x)$ is the
associated Legendre function of the second kind \cite{Abra}. From $r\gg
r_{H} $ we get the constraint on the ratio $r/a$ in (\ref{jlAdS}): $r/a\gg
\left( r_{G}/a\right) ^{1-2/D}$. If the length scales $r_{G}$ and $a$ are of
the same order of magnitude, one has $r/a\gg 1$ and, by using the large
argument asymptotic for the function $q_{\alpha }^{\mu }(x)$, from (\ref%
{jlAdS}) we get
\begin{equation}
\langle j^{l}\rangle \approx \frac{4ea^{-D}\Gamma \left( \nu +D/2+1\right) }{%
\pi ^{D/2-1}\Gamma \left( \nu +1\right) \left( 2\pi r/a\right) ^{D+2\nu +2}}%
\sideset{}{'}{\sum}_{\mathbf{n}_{q}}\frac{n_{l}\sin \left( \mathbf{n}%
_{q}\cdot \boldsymbol{\tilde{\alpha}}\right) }{(n_{p+2}^{2}+\cdots
+n_{D}^{2})^{\frac{D}{2}+\nu +1}},  \label{jlAdS2}
\end{equation}%
where the prime on the summation sign means that the term with $%
n_{p+2}=\cdots =n_{D}=0$ should be excluded. For $r_{G}\ll a$ the ratio $r/a$
can be of the order of one and more general expression (\ref{jlAdS}) has to
be used.

The influence of branes on the vacuum current density for a charged scalar
field in locally AdS spacetime with a toroidal subspace has been
investigated in \cite{Bell15AdS,Bell16AdS}. Branes parallel to the AdS
boundary have been considered and the corresponding formulas approximate the
large-distance asymptotic for the current density around a cylindrical black
hole in the presence of additional boundaries with $r=\mathrm{const}$
(spherical branes).

\section{Conclusion}

\label{sec:Conc}

We have investigated the vacuum currents for a charged scalar field in
Rindler spacetime with a toroidally compact subspace. It is assumed that the
field is prepared in the FR vacuum state and obeys quasi-periodicity
condition (\ref{PC}) along $l$th compact dimension. For an external gauge
field $A_{\mu }$ a simple configuration is considered with constant
components in the compact subspace. Those components and the phases in the
condition (\ref{PC}) appear in the expressions for the VEVs of physical
quantities in the form of combination (\ref{alftilde}). The latter is
interpreted in terms of the magnetic flux threading the compact dimension.
The complete set of scalar mode functions realizing the FR vacuum is given
by (\ref{fisigma}). The components of the momentum in the compact subspace
are quantized by the periodicity conditions and the corresponding
eigenvalues are given by (\ref{kltild}). We have started the investigation
by evaluating the mode sum for the Hadamard function. The latter is
presented in the form (\ref{Gdif2}), where $G_{\mathrm{M}}(x,x^{\prime })$
is the corresponding function for the Minkowski vacuum in flat spacetime
with spatial topology $R^{p+1}\times (S^{1})^{q}$ and the last term
describes the difference in the correlations of the vacuum fluctuations in
the FR and Minkowski vacua.

The VEV of the current density is obtained from the Hadamard function by
using the formula (\ref{jW}). The charge and current densities along
uncompact dimensions vanish. The current density along the $l$th compact
dimension is presented in two equivalent forms, given by (\ref{jr}) and (\ref%
{jr2}). In both the representations the corresponding current density in the
Minkowski vacuum is explicitly extracted. The latter is investigated in the
literature and the corresponding VEV is expressed as (\ref{jrM}) or,
equivalently, by (\ref{jrM2}). The projection of the current density on $l$%
th compact dimension is an odd periodic function of the parameter $\tilde{%
\alpha}_{l}$ and an even periodic function of the parameters $\tilde{\alpha}%
_{i}$, $i\neq l$, with period $2\pi $. In terms of the magnetic flux
enclosed by compact dimensions, this corresponds to periodicity with period
equal to the flux quantum. We have shown that the current density vanishes
on the Rindler horizon that corresponds to the limit $\xi \rightarrow 0$.
The large values of $\xi $ correspond to small accelerations and the
difference in the current densities for the FR and Minkowski vacua is
exponentially small (see (\ref{jrlarge})). For small values of the length of
compact dimension the difference of the current densities between the FR and
Minkowski vacua is exponentially suppressed and the separate current
densities along the $l$th compact dimension behave like $1/L_{l}^{D}$. For
large values of $L_{l}$ and for $m=0$, $\tilde{\alpha}_{i}=0$, $i\neq l$,
one has power-law decay of the Minkowskian contribution, $\langle
j^{l}\rangle _{\mathrm{M}}\propto 1/L_{l}^{p+1}$. For $\omega _{0l}\neq 0$
the large $L_{l}$ asymptotic of the VEV $\langle j^{l}\rangle _{\mathrm{M}}$
contains an exponential factor $e^{-L_{l}\omega _{0l}}$ and it falls off
more rapidly. In the large $L_{l}$ limit the decay of the difference $%
\langle j^{l}\rangle -\langle j^{l}\rangle _{\mathrm{M}}$ is stronger than
that for $\langle j^{l}\rangle _{\mathrm{M}}$.

As an application of the obtained results we have considered the
near-horizon vacuum currents around cylindrical black holes. The
corresponding exterior geometry is described by the line element (\ref{ds2bh}%
). Near the horizon it is approximated by the Rindler-like metric (\ref{ds2}%
) considered in Section \ref{sec:Modes} with the lengths of the compact
dimensions $L_{l}=2\pi r_{H}$ for all $l=p+2,\ldots ,D$. At large distances
from the horizon the geometry of cylindrical black holes is approximated by
a locally AdS spacetime with a toroidally compact subspace. The lengths of
the corresponding compact dimensions are expressed in terms of the AdS
curvature scale $a$ as $L_{l}=2\pi a$. Similar to the near-horizon
approximation, all the compact dimensions have the same length. The vacuum
currents in that geometry have been investigated in \cite{Beze15} and are
given by the right-hand side of (\ref{jlAdS}).

In the discussion above, as a local characteristic of the FR vacuum state
the expectation value of the current density is considered. By using the
Hadamard function (\ref{Gdif2}), we can investigate the expectation values
for other physical characteristics bilinear in the field operator. Among
those characteristics, an important quantity is the VEV of the
energy-momentum tensor and the corresponding results for the FR vacuum will
be presented elsewhere. Here we mention that the influence of nontrivial
spatial topology on the energy of the Minkowski vacuum has been widely
considered in the literature for scalar, fermionic and vector fields (see,
for example, \cite{Khan14},\cite{Eder08}-\cite{Hari21} and references
therein).

\section*{Acknowledgments}

A.A.S. was supported by the grants No. 20RF-059 and No. 21AG-1C047 of the Committee of Science
of the Ministry of Education, Science, Culture and Sport RA.

\appendix

\section{Representation for the Minkowskian Hadamard function}

\label{sec:App2}

In this section we provide a representation for the Minkowskian Hadamard
function that is convenient in order to compare the topological effects in
Minkowski and Rindler spacetimes with compact dimensions. We consider a
locally Minkowskian spacetime with the line element (\ref{ds2M}), where the
subspace $(x^{2},x^{3},\ldots ,x^{D})$ has the topology $R^{p}\times S^{q}$.
By using the corresponding mode sum formula and the Minkowskian
eigenfunctions it can be seen that the Minkowskian Hadamard function may be
presented in the form
\begin{equation}
G_{\mathrm{M}}(x,x^{\prime })=\frac{2^{-p}}{\pi ^{p+1}V_{q}}\int d\mathbf{k}%
_{p}\sum_{\mathbf{n}_{q}}e^{i\mathbf{k}\cdot \Delta {\mathbf{x}}}K_{0}\left(
\lambda _{\mathbf{k}}\sqrt{\xi ^{2}+\xi ^{\prime 2}-2\xi \xi ^{\prime }\cosh
\Delta \tau }\right) ,  \label{Gmin}
\end{equation}%
where $\xi ^{2}+\xi ^{\prime 2}-2\xi \xi ^{\prime }\cosh \Delta \tau =(x_{%
\mathrm{M}}^{1}-x_{\mathrm{M}}^{\prime 1})^{2}-(t_{\mathrm{M}}-t_{\mathrm{M}%
}^{\prime })^{2}$. For the further transformation we use the integral
representation \cite{Prud2}
\begin{equation}
K_{0}\left( \lambda \sqrt{\xi ^{2}+\xi ^{\prime 2}+2\xi \xi ^{\prime }\cos b}%
\right) =\frac{2}{\pi }\int_{0}^{\infty }d\omega \,\cosh (b\omega
)K_{i\omega }(\lambda \xi )K_{i\omega }(\lambda \xi ^{\prime }),
\label{K0KK}
\end{equation}%
Taking $b=\pi \pm i\Delta \tau $ we get%
\begin{equation}
K_{0}\left( \lambda \sqrt{\xi ^{2}+\xi ^{\prime 2}+2\xi \xi ^{\prime }\cos
\left( \pi \pm i\Delta \tau \right) }\right) =\frac{2}{\pi }\int_{0}^{\infty
}d\omega \,\cosh \left[ \left( \pi \pm i\Delta \tau \right) \omega \right]
K_{i\omega }(\lambda \xi )K_{i\omega }(\lambda \xi ^{\prime }).
\label{K0KK1}
\end{equation}%
From here it follows that%
\begin{equation}
K_{0}\left( \lambda \sqrt{\xi ^{2}+\xi ^{\prime 2}-2\xi \xi ^{\prime }\cosh
\Delta \tau }\right) =\frac{2}{\pi }\int_{0}^{\infty }d\omega \,\cosh \left(
\pi \omega \right) \cos \left( \omega \Delta \tau \right) K_{i\omega
}(\lambda \xi )K_{i\omega }(\lambda \xi ^{\prime }).  \label{K0KK2}
\end{equation}%
Combining this relation with (\ref{Gmin}), the Minkowskian Hadamard function
is presented in the form (\ref{Gmin2}).

\section{Rindler spacetime with compact dimensions as a near-horizon
geometry for cylindrical black holes}

\label{sec:App1}

In this section we consider the near-horizon limit of the exterior geometry
for a cylindrical black hole given by (\ref{ds2bh}). We write $r=r_{H}\left(
1+x/D\right) $ and expand the metric for small $x$. To the leading order one
gets $f(r)\approx Mx/r_{H}^{D-2}$ and
\begin{equation}
ds_{\mathrm{bh}}^{2}\approx \frac{Mx}{r_{H}^{D-2}}dt^{2}-\frac{a^{2}dx^{2}}{%
D^{2}x}-r_{H}^{2}\sum_{i=1}^{p}\left( dy^{i}\right)
^{2}-r_{H}^{2}\sum_{i=1}^{q}\left( d\phi _{i}\right) ^{2}.  \label{ds2bh3}
\end{equation}%
Introducing the coordinate $\xi $ in accordance with $x=\left( D\xi
/2a\right) ^{2}$, the line element is rewritten as%
\begin{equation}
ds_{\mathrm{bh}}^{2}\approx \frac{D^{2}M}{4a^{2}r_{H}^{D-2}}\xi
^{2}dt^{2}-d\xi ^{2}-r_{H}^{2}\sum_{i=1}^{p}\left( dy^{i}\right)
^{2}-r_{H}^{2}\sum_{i=1}^{q}\left( d\phi _{i}\right) ^{2}.  \label{ds2bh4}
\end{equation}%
In terms of the new coordinates%
\begin{equation}
\tau =\frac{D\sqrt{M}}{2ar_{H}^{D/2-1}}t,\;x^{i+1}=r_{H}y^{i},%
\;x^{p+l+1}=r_{H}\phi _{l},  \label{tau}
\end{equation}%
with $i=1,\ldots ,p$, $l=1,\ldots ,q$, this line element attains a locally
Rindler form%
\begin{equation}
ds_{\mathrm{bh}}^{2}\approx \xi ^{2}d\tau ^{2}-d\xi
^{2}-\sum_{i=2}^{D}\left( dx^{i}\right) ^{2},  \label{ds2bh5}
\end{equation}%
where $-\infty <x^{i}<+\infty $ for $i=2,\ldots ,p+1$, and $0\leq x^{i}\leq
2\pi r_{H}$ for $i=p+2,\ldots ,D$.


\begin{thebibliography}{99}
\bibitem{Birr82} N. D. Birrell and P. C. W. Davies, \textit{Quantum Fields
in Curved Space} (Cambridge University Press, Cambridge, 1982).

\bibitem{Full73} S. A. Fulling, Nonuniqueness of canonical field
quantization in Riemannian space-time. Phys. Rev. D \textbf{7}, 2850 (1973).

\bibitem{Empa99} R. Emparan, AdS/CFT duals of topological black holes and
the entropy of zero-energy states. J. Hihg Energy Phys. 06(1999)036.

\bibitem{Hami06} A. Hamilton, D.N. Kabat, G. Lifschytz, and D.A. Lowe,
Holographic representation of local bulk operators. Phys. Rev. D \textbf{74}%
, 066009 (2006).

\bibitem{Pari12} M. Parikh, P. Samantray, and E. Verlinde, Rotating
Rindler-AdS space. Phys. Rev. D \textbf{86}, 024005 (2012).

\bibitem{Czec12} B. Czech, J. L. Karczmarek, F. Nogueira, and M. Van
Raamsdonk, Rindler quantum gravity. Class. Quant. Grav. \textbf{29}, 235025
(2012).

\bibitem{Sama14} P. Samantray and T. Padmanabhan, Conformal symmetry,
Rindler space, and the AdS/CFT correspondence. Phys. Rev. D \textbf{90},
047502 (2014).

\bibitem{Pari18} M. Parikh and P. Samantray, Rindler-AdS/CFT. J. Hihg Energy
Phys. 10(2018)129.

\bibitem{Most97} V. M. Mostepanenko and N. N. Trunov, \textit{The Casimir
Effect and Its Applications} (Clarendon,Oxford, 1997).

\bibitem{Milt02} K. A. Milton, \textit{The Casimir Effect: Physical
Manifestation of Zero-Point Energy} (World Scientific, Singapore, 2002).

\bibitem{Bord09} M. Bordag, G. L. Klimchitskaya, U. Mohideen, and V. M.
Mostepanenko, \textit{Advances in the Casimir Effect} (Oxford University
Press, New York, 2009).

\bibitem{Casi11} Casimir Physics, edited by D. Dalvit, P. Milonni, D.
Roberts, and F. da Rosa, \textit{Lecture Notes in Physics} Vol. 834
(Springer-Verlag, Berlin, 2011).

\bibitem{Cand77} P. Candelas and D. Deutsch, Proc. R. Soc. London A \textbf{%
354}, 79 (1977).

\bibitem{Saha02} A. A. Saharian, Polarization of the Fulling-Rindler vacuum
by a uniformly accelerated mirror. Class. Quantum Grav. \textbf{19}, 5039
(2002).

\bibitem{Avag02} R. M. Avagyan, A. A. Saharian, and A. H. Yeranyan, Casimir
effect in the Fulling-Rindler vacuum. Phys. Rev. D \textbf{66}, 085023
(2002).

\bibitem{Saha04} A. A. Saharian, R. S. Davtyan, and A. H. Yeranyan, Casimir
energy in the Fulling-Rindler vacuum. Phys. Rev. D \textbf{69}, 085002
(2004).

\bibitem{Saha04b} A. A. Saharian and M. R. Setare, Surface vacuum energy and
stresses on a plate uniformly accelerated through the Fulling-Rindler
vacuum. Class. Quantum Grav. \textbf{21}, 5261 (2004).

\bibitem{Saha04c} A. A. Saharian and M. R. Setare, Casimir energy-momentum
tensor for a brane in de Sitter spacetime. Phys. Lett. B \textbf{584}, 306
(2004).

\bibitem{Saha06} A. A. Saharian, R. M. Avagyan, and R. S. Davtyan, Wightman
function and Casimir densities for Robin plates in the Fulling-Rindler
vacuum. Int. J. Mod. Phys. A \textbf{21}, 2353 (2006).

\bibitem{Saha05} A. A. Saharian and M. R. Setare, Casimir densities for a
spherical brane in Rindler-like spacetimes. Nucl. Phys. B \textbf{724}, 406
(2005).

\bibitem{Saha06b} A. A. Saharian and M. R. Setare, Surface Casimir densities
on a spherical brane in Rindler-like spacetimes Phys. Lett. B \textbf{637},
5 (2006).

\bibitem{Lemo95} J. P. S. Lemos, Cylindrical black hole in general
relativity. Phys. Lett. B \textbf{353}, 46 (1995).

\bibitem{Lemo96} J. P. S. Lemos and V. T. Zanchin, Rotating charged black
strings and three-dimensional black holes. Phys. Rev. D \textbf{54}, 3840
(1996).

\bibitem{Cai96} R.-G. Cai and Y.-Z. Zhang, Black plane solutions in
four-dimensional spacetimes. Phys. Rev. D \textbf{54}, 4891 (1996).

\bibitem{Horo98} G. T. Horowitz and R. C. Myers, AdS-CFT correspondence and
a new positive energy conjecture for general relativity. Phys. Rev. D
\textbf{59}, 026005 (1998).

\bibitem{Mart00} C. Martinez, C. Teitelboim, and J. Zanelli, Charged
rotating black hole in three space-time dimensions. Phys. Rev. D \textbf{61}%
, 104013 (2000).

\bibitem{Awad03} A. M. Awad, Higher dimensional charged rotating solutions
in (A)dS space-times. Class. Quant. Grav. \textbf{20}, 2827 (2003).

\bibitem{Gaet17} M. B. Gaete, L. Guajardo, and M. Hassa\"{\i}ne, A
Cardy-like formula for rotating black holes with planar horizon. J. Hihg
Energy Phys. 04(2017)092.

\bibitem{Cris08} L. C. B. Crispino, A. Higuchi, and G. E. A. Matsas, The
Unruh effect and its applications. Rev. Mod. Phys. \textbf{80}, 787 (2008).

\bibitem{Abra} \textit{Handbook of Mathematical Functions}, edited by M.
Abramowitz, I. A. Stegun (Dover, New York, 1972).

\bibitem{Cand76} P. Candelas and D. J. Raine, Quantum field theory on
incomplete manifolds. J. Math. Phys. \textbf{17}, 2101 (1976).

\bibitem{Khan14} F. C. Khanna, A. P. C. Malbouisson, J. M. C. Malbouisson,
and A. E. Santana, Quantum field theory on toroidal topology: Algebraic
structure and applications. Phys. Rep. \textbf{539}, 135 (2014).

\bibitem{Raje20} K. Rajeev and T. Padmanabhan, Exploring the Rindler vacuum
and the Euclidean plane. J. Math. Phys. \textbf{61}, 062302 (2020).

\bibitem{Wats66} G. N. Watson, A Treatise on the Theory of Bessel Functions
(Cambridge University Press, Cambridge, England, 1966).

\bibitem{Beze13} E. R. Bezerra de Mello and A. A. Saharian, Finite
temperature current densities and Bose-Einstein condensation in
topologically nontrivial spaces. Phys. Rev. D \textbf{87}, 045015 (2013).

\bibitem{Bene98} A. DeBenedictis, Gravitational effects of quantum fields in
the interior of a cylindrical black hole. Class. Quantum Grav. \textbf{16},
1955 (1999).

\bibitem{Bene99} A. DeBenedictis, $\langle \phi ^{2}\rangle $ in the
spacetime of a cylindrical black hole. Gen. Rel. Grav. \textbf{32}, 1549
(1999).

\bibitem{Beze15} E. R. Bezerra de Mello, A. A. Saharian, and V. Vardanyan,
Induced vacuum currents in anti-de Sitter space with toral dimensions. Phys.
Lett. B \textbf{741}, 155 (2015).

\bibitem{Bell15AdS} S. Bellucci, A. A. Saharian, and V. Vardanyan, Vacuum
currents in braneworlds on AdS bulk with compact dimensions. J. Hihg Energy
Phys. 11(2015)092.

\bibitem{Bell16AdS} S. Bellucci, A. A. Saharian, and V. Vardanyan, Hadamard
function and the vacuum currents in braneworlds with compact dimensions:
Two-brane geometry. Phys. Rev. D \textbf{93}, 084011 (2016).

\bibitem{Prud2} A. P. Prudnikov, Yu. A. Brychkov, and O. I. Marichev,
Integrals and Series (Gordon and Breach, New York, 1986), Vol. 2.

\bibitem{Eder08} A. Edery and V. N. Marachevsky, Compact dimensions and the
Casimir effect: the Proca connection. J. High Energy Phys. 12(2008)035.

\bibitem{Eliz09} E. Elizalde, S. D. Odintsov, and A. A. Saharian, Repulsive
Casimir effect from extra dimensions and Robin boundary conditions: From
branes to pistons. Phys. Rev. D \textbf{79}, 065023 (2009).

\bibitem{Bell09} S. Bellucci1 and A. A. Saharian, Fermionic Casimir
densities in toroidally compactified spacetimes with applications to
nanotubes. Phys. Rev. D \textbf{79}, 085019 (2009).

\bibitem{Teo11} L. P. Teo, Electromagnetic Casimir piston in
higher-dimensional spacetimes. Phys. Rev. D \textbf{83}, 105020 (2011).

\bibitem{Cao13} C. J. Cao, M. van Caspel, and A. R. Zhitnitsky, Topological
Casimir effect in Maxwell electrodynamics on a compact manifold. Phys. Rev.
D \textbf{87}, 105012 (2013).

\bibitem{Hari21} S. R. Haridev and P. Samantray, Revisiting vacuum energy in
compact spacetimes. arXiv:2106.12171.
\end{thebibliography}
\end{document}